# A Derivation Of The Scalar Propagator In A Planar Model In Curved Space

S. G. Kamath*

Department of Mathematics, Indian Institute of Technology Madras, Chennai 600 036, India
*e-mail: kamath@iitm.ac.in

**Abstract**: Given that the free massive scalar propagator in 2 + 1 dimensional Euclidean space is $D(x-y)=\frac{1}{4\pi\rho}e^{-m\rho}$ with $\rho^2=(x-y)^2$ we present the counterpart of $D(x-y)$ in curved space with a suitably modified version of the Antonsen - Bormann method instead of the familiar Schwinger - de Witt proper time approach, the metric being defined by the rotating solution of Deser et al. of the Einstein field equations associated with a single massless spinning particle located at the origin.

**Keywords**: massive scalar field theory, 2 + 1 dimensions, curved space - time, stationary solutions, heat kernel, propagator.

**PACS** : 04.60Kz, 04.62+v, 11.10Kk

## INTRODUCTION

This is the second of two papers associated with the Lagrangian density

$$L=\frac{1}{2}g^{\mu\rho}\,\partial_\mu\varphi\partial_\rho\varphi-\frac{1}{2}m^2\varphi^2 \tag{1}$$

in 2 + 1 dimensions; it extends the scope of the first [1] with this objective: Derive in curved space the counterpart of the propagator $\langle 0|T^*(\varphi(x)\varphi(y)|0\rangle$ that is given in Euclidean flat space – with $g^{\alpha\beta}=diag(1,1,1)$ – by $D(x-y)=\frac{1}{4\pi\rho}e^{-m\rho}$ with $\rho^2=(x-y)^2$. This report is inspired by two recent papers of the Napoli group of Bimonte et al.[2,3] and is based on the method proposed by Antonsen and Bormann[4,5]. To elaborate, we recall here that Bimonte et al.[2] obtain the Green's functions in curved space using the Fock – Schwinger - de Witt method[6,7,8], an important feature of which is that it is valid only in the asymptotic region – see eq.(2.4) in Ref.2 et seq.. Again, the use of the Fermi normal coordinates in Bunch and Parker[9] leads to a momentum space representation for the Feynman propagator in curved space but the said infirmity that obtains in [2] still prevails, while the recent calculation[3] of the energy momentum tensor for a Casimir apparatus but in a weak gravitational field using a Fermi coordinate system only reinforces our aim. In this connection, the work of Antonsen and Bormann [4, 5] holds some promise but with a modification that is imperative for our purposes, given that in Ref.4 the last term in eq.(23) has been set to zero[10] thus undermining the utility of their method[4,5].

## The Antonsen – Bormann Method

We begin with the operator $B \equiv -\partial^{\mu}\left(g_{\mu\rho}\,\partial^{\rho}\right) - m^2$ obtained from eq.(1) above and work with the heat kernel $G(x,x';\sigma)$ obeying

$$BG(x,x';\sigma) = -\frac{\partial G}{\partial \sigma} \tag{2}$$

with $G(x,x';\sigma \to 0) = \delta^{(3)}(x-x')$ .The vierbeins $e^{\alpha}_{\beta}$ defined by $g^{\alpha\beta} = \eta^{\mu\nu}\, e^{\alpha}_{\mu} e^{\beta}_{\nu}$ with $\eta^{\mu\nu} = \mathrm{diag}(1,-1,-1)$ can now be used to rework B into the form

$$B = -\eta^{ab}\,\partial_a \partial_b - m^2 - e^{m}_{\alpha}\,\partial_m\,(e^{\alpha}_{a})\partial^{a} \tag{3}$$

Writing $G(x,x';\sigma)$ in terms of the flat space heat kernel $G_0(x,x';\sigma)$ as

$$G(x,x';\sigma) = G_0 e^{-\frac{1}{2}\int e^{m}_{\alpha}\,\partial_m\,(e^{\alpha}_{n})dx^{n}}\, e^{-T} \tag{4}$$

with $(-\eta^{ab}\,\partial_a\partial_b - m^2)G_0 = -\frac{\partial G_0}{\partial\sigma}$ and $G_0(x,x';\sigma \to 0) = \delta^{(3)}(x-x')$ enables us to remove the last term in (3) and obtain as in Ref.4

$$\partial^{a}\partial_{a}T - \partial^{a}T\partial_{a}T + f + 2\,\frac{\partial^{\mu} G_0}{G_0}\,\partial_{\mu}T = \frac{\partial T}{\partial\sigma} \tag{5}$$

with $f = \frac{1}{4}(e^{m}_{\alpha}\,\partial(e^{\alpha}_{b}))^2 + \frac{1}{2}\partial^{n}\left(e^{m}_{\beta}\,\partial_m\,(e^{\beta}_{n})\right)$. The last term in (5) can be reworked as $-\frac{(x-x')^{a}}{\sigma}\,\partial_a T$ recalling that in Euclidean space $G_0 = (4\pi\sigma)^{-3/2} e^{-\frac{(x-x')^2}{4\sigma} - m^2\sigma}$. We shall now write T as

$$T = \frac{\tau_{-1}}{\sigma} + \sum_{k=0}^{\infty} \tau_k \sigma^k$$

and not as in eq.(24) in Ref.4 to meet the twin requirements of : a. the boundary condition on the heat kernels G and $G_0$ as $\sigma \to 0$ thus setting $\tau_0 = -\frac{1}{2}\int e^{n}_{\alpha}\,\partial_n\,(e^{\alpha}_{m})dx^{m}$ and b. retaining the extra term $-\frac{(x-x')^{a}}{\sigma}\,\partial_a T$. Repeating the steps in Ref.4 one now gets $\tau_{-1}$ to be a solution of

$$\partial^{a}\partial_{a}\tau_{-1} - 2\partial^{a}\tau_0\partial_a\tau_{-1} - (x-x')^{a}\partial_a\tau_0 = 0 \tag{6}$$

Besides eq.(6) here are a few more entries that are the counterparts of those given by eq.(14) in Ref.5:

$$2\partial^{a}\tau_{-1}\partial_a\tau_1 + (x-x')^{a}\partial_a\tau_1 - f + \tau_1 + \partial^{a}\tau_0\partial_a\tau_0 - \partial^{a}\partial_a\tau_0 = 0 \tag{7}$$

$$\partial^{a}\partial_a\tau_1 - 2\partial_a\tau_{-1}\partial^{a}\tau_2 - 2\partial_a\tau_1\partial^{a}\tau_0 - (x-x')^{a}\partial_a\tau_2 = 2\tau_2 \tag{8}$$

## Some Examples of the Vierbeins $e_a^\mu$ and $e_\mu^a$

Adopting the stationary solutions of the Einstein field equations given in Deser , Jackiw and 'tHooft[11] and Clement[12] wherein

$$g^{00} = 1 - \frac{\lambda^2}{r^2}, g^{01} = -\frac{\lambda y}{r^2}, g^{02} = \frac{\lambda x}{r^2}, g^{11} = -1, g^{12} = 0, g^{22} = -1 \qquad (9)$$

with $r = |\boldsymbol{r}|$, it is easy to determine the vierbeins $e_a^\mu$; their counterparts $e_\mu^a$ are got from $g_{\alpha\beta} = \eta_{\mu\nu} e_\alpha^\mu e_\beta^\nu$ where $g_{\alpha\beta}$ is given by[11,12]

$$g_{00} = 1, g_{01} = -\frac{\lambda y}{r^2}, g_{02} = \frac{\lambda x}{r^2}, g_{11} = -1 + \left(\frac{\lambda y}{r^2}\right)^2, \ g_{12} = -\left(\frac{\lambda^2 xy}{r^4}\right), g_{22} = -1 + \left(\frac{\lambda x}{r^2}\right)^2 \qquad (10)$$

with $2\pi\lambda = \kappa \mathrm{J}$ , $\kappa = 8\pi G$ being the gravitational constant and $\mathrm{J} = |\boldsymbol{J}|$ being the spin of the massless particle located at the origin(see eq.(20) in Clement[12]); note that $det g_{\mu\nu} = 1$ so the action $S = \int d^3x\, L$ with $L$ as in (1). As an example for which $\tau_0 = 0$ and $f = 0$ we have:

$$e_a^\mu : e_0^0 = \frac{i\lambda y}{r^2}, e_0^1 = i, e_0^2 = 0; e_1^0 = i, e_1^1 = 0, e_1^2 = 0; e_2^0 = \frac{\lambda x}{r^2}, e_2^1 = 0, e_2^2 = -1$$

$$e_\mu^a : e_0^0 = 1, e_0^1 = 0 = e_0^2; e_1^0 = -\frac{\lambda y}{r^2}, e_1^1 = \frac{1}{\sqrt{2}}, e_1^2 = -\frac{1}{\sqrt{2}}; \ e_2^0 = \frac{\lambda x}{r^2}, e_2^1 = -\frac{1}{\sqrt{2}}, e_2^2 = -\frac{1}{\sqrt{2}}$$

An example which yields $\tau_0 = \frac{i\lambda x}{2r^2}$ and $f = \frac{\lambda^2}{4r^4}$ is :

$$e_a^\mu : e_0^0 = 1, e_0^1 = 0, e_0^2 = 0; e_1^0 = \frac{\lambda x}{r^2}, e_1^1 = 0, e_1^2 = -1; e_2^0 = \frac{\lambda y}{r^2}, e_2^1 = 1, e_2^2 = 0$$

$$e_\mu^a : \ e_0^0 = 0, e_0^1 = -i, e_0^2 = 0; e_1^0 = \frac{i}{\sqrt{2}}, e_1^1 = \frac{i\lambda y}{r^2}, e_1^2 = \frac{1}{\sqrt{2}}; e_2^0 = \frac{i}{\sqrt{2}}, e_2^1 = -\frac{i\lambda x}{r^2}, e_2^2 = -\frac{1}{\sqrt{2}}$$

It is easy to check for this example that $\partial^a \partial_a \tau_0 = 0$ and that $\partial_a \tau_0 \partial^a \tau_0 = \frac{\lambda^2}{4r^4}$ , which is $f$ above. Clearly, it is not difficult to cite other examples involving for instance the circular functions[13].

## Solving Eqs.(6) – (8) Above

With $\tau_{-1} = \chi_{-1} e^{\tau_0}$ one gets from (6)

$$\partial^a \partial_a \chi_{-1} + (\partial^a \partial_a \tau_0 - \ 2\, \partial_a \tau_0 \partial^a \tau_0)\chi_{-1} = (x - x')^a \partial_a \tau_0 e^{-\tau_0} \qquad (11)$$

the solution to which can be attempted using a Green's function, thus yielding $\tau_{-1}$.The learned reader will notice that with the coefficient of $\chi_{-1}$ above set at $\frac{\lambda^2}{2r^4}$, (11) now becomes an inhomogeneous Helmholtz equation[14] in Minkowski space. With $f$ and $\tau_{-1}$ known the method of Lagrange readily yields a solution to (7), it being a first order partial differential equation; thus $\tau_1$ is obtained.

$\tau_2$ can now be got from(8) since the latter is also a first order pde. This exercise can now be repeated to obtain the $\tau_j$ , $j \geq 3$ resulting in the full form of T above and thus the heat kernel $G(x, x^{'}; \sigma)$, whose integral – see eq.(5) in Ref.5 – yields the required propagator that is the counterpart of $D(x-y)$ in curved space.

# ACKNOWLEDGEMENTS

I thank K.P.Deepesh for his generous assistance in the preparation of this manuscript and am grateful to the Indian Institute of Technology Madras for their financial support.